%% file: main.tex
\documentclass{article} 
\usepackage{ml4se2019,times}

\input{math_commands.tex}

\usepackage{booktabs}
\usepackage{url}
\usepackage{subfig}
\usepackage{caption}
\usepackage{graphicx}
\usepackage{svg}
\usepackage{amsmath}
\usepackage[english]{babel}
\usepackage[autolanguage]{numprint}

\newcommand{\Gitlab}{GitLab}
\newcommand{\Hercules}{Hercules}
\newcommand{\Gitbase}{Gitbase}

\title{Identifying collaborators in large codebases}

\author{Waren Long, Vadim Markovtsev, Hugo Mougard, Egor Bulychev \& Jan Hula\\
source\{d\}\\
Madrid, Spain \\
\texttt{\{waren,vadim,hugo,egor,jan\}@sourced.tech}
}

\mlsefinalcopy 
\begin{document}

\maketitle

\begin{abstract}
The way developers collaborate inside and particularly across teams often escapes management's attention, despite a formal organization with designated teams being defined.
Observability of the actual, organically formed engineering structure provides decision makers invaluable additional tools to manage their talent pool.

To identify existing inter and intra-team interactions---and suggest relevant opportunities for suitable collaborations---this paper studies contributors' commit activity, usage of programming languages, and code identifier topics by embedding and clustering them.
We evaluate our findings collaborating with the \Gitlab{} organization, analyzing 117 of their open source projects. We show that we are able to restore their engineering organization in broad strokes, and also reveal hidden coding collaborations as well as justify in-house technical decisions.
\end{abstract}

\section{Introduction}
\label{sec:introduction}

Many organizations undergo rapid digital transformation nowadays.
In such a setting, a common challenge is the understanding of software development dynamics.
Horizontal relations between development teams are typically obscure and unobvious at a glance.
Yet understanding them is important to manage organizational structure, plan hiring, and prioritize work.

This paper introduces complementary methods to analyze contributors to open source projects. We study several orthogonal contributor features to obtain an overview of existing collaborations and evaluate our methods on the \Gitlab{} organization, a large software company with over 500 employees and 110 projects with more than 2,000 external contributors as of 2019.
The considered features include the alignment of commit time series, the developer experience reflected in contributed lines of code (LoC), and topic models related to code naturalness~\citep{Allamanis2017Naturalness}. We harness dimensionality reduction and clustering for visualization.

\section{Related Work}
\label{sec:related-work}
Some researchers have already explored developer collaboration and recommendation through analysis of the coding activities.
For example, \cite{Montandon2019IdentifyingEI} made use of supervised machine learning classifiers to identify experts in three popular JavaScript libraries.
However, that approach depends on manually labeled classes, so it may be difficult to generalize.
Along the same line, \cite{Greene2016CVExplorerIC} have developed a tool to extract and visualize developers' skills by using both keywords from the project's README file and commit messages combined with their portfolio metadata, which they collected from third-party Community Question Answering websites.
\cite{Hauff2015MatchingGD} had a different objective to directly match developers to relevant jobs by comparing their embeddings. Those embeddings were generated from the query describing the job and the README file of the project, but they do not leverage the source code itself.
Regarding the topic modeling technique, it has been applied on source code identifiers by \cite{Markovtsev2017TM}, but without analyzing individual contributions or focusing on potential coding collaborations.

Differing from these, we detect collaborators by working with fine-grained information from every commit: we analyze every file in the repository's history.

\section{Methodology}
\label{sec:methodology}

This section presents a combination of three complementary developer analyses, and each clarifies a particular aspect of developers' characteristics.

\subsection{Data Processing}
\label{sec:data-processing}

To gain the quantitative insight into source code required by further analyses, we employ \Hercules{}\footnote{\url{https://github.com/src-d/hercules}} and \Gitbase{}\footnote{\url{https://github.com/src-d/gitbase}}---tools that allow efficient and rapid mining of Git repositories.
We gather the number of contributed lines per programming language per developer, the ownership information about each file (who edited each line the last), and the commit dates.
We also extract the identifiers from each file and store their frequencies.
There is the need for matching Git signatures to contributors. Therefore we aggregate all co-occurring emails and names to a single identity by finding connected components in the corresponding undirected co-occurrence graph. We have to exclude signature stubs such as \texttt{gitlab@localhost}. The help of GitLab API is intentionally avoided for versatility reasons.

\subsection{Commit Time Series}
\label{sec:commit-time-series}

The first analysis is grounded on aligning developers' commits time series.
People who work in a team tend to create commits at correlated times, because they often work on related, mutually dependent features.
The relative commit frequency depends on the shared team planning, notably the deadlines.
We found empirically that the absolute values of the time series are less important than the distribution shape.

Thus we extract how many commits were done by each developer per day, divide the values by the mean to normalize the resulting time series, and calculate the pairwise distance matrix using Fast Dynamic Time Warping~\citep{Salvador2007TAD}.
Then we perform DBSCAN~\citep{Ester1996DAD} clustering and run UMAP~\citep{McInnes2018UMAP} dimensionality reduction to visualize the clusters. Both mentioned algorithms operate on the explicit distance matrix.

\subsection{Contributions by Programming Languages}
\label{sec:contributions-by-programming-languages}

The following way of grouping developers is based on the programming language, the amount and the nature of contributions (line additions, modifications, and deletions).
Intuitively, these coarsely correlate with developer seniority and activity domains. Hence we exclude text markup files from the analysis.
We reduce the influence of outliers by saturating all the values to the 95\textsuperscript{th} percentile.
Our distance metric is L\textsuperscript{2}.
We visualize the distribution by applying UMAP, which preserves the local topological relations that should be highlighted in our case.
We calculate clusters in the embedding space using K-Means, and their number is determined by the elbow rule.
Running K-Means in the original space produces worse results due to its high-dimensionality and sparsity.

\subsection{Source Code Identifiers}
\label{sec:source-code-identifiers}

To complement the commit activity and the programming language usage analyses, we leverage the content of developers' contributions in order to assign them topics that are representative of their activity.
We compute the representation for each code file based on its content as a bag of TF-IDF scores of its identifiers.
We then represent each developer by aggregating the representations of the code files they edited with a sum weighted by the proportion of the owned lines in each file. Finally, we apply ARTM~\citep{Vorontsov2015BigARTM} to find decorrelated, sparse topics fitted to those representations and label them manually having obtained their most likely and specific terms.

\section{Results}
\label{sec:results}

We evaluate the three presented analyses on the codebase of the \Gitlab{} organization.
That comprised \numprint{117} repositories, totaling \numprint{145} programming languages, \numprint{44272} files and \numprint{12895956} lines of code in April 2019.
Among the \numprint{2956} contributors, we detected \numprint{210} current \Gitlab{} employees split in 37 teams (this is a lower bound given the algorithmic difficulties detailed in Section~\ref{sec:data-processing}). A mapping from developers to teams is publicly available\footnote{https://gitlab.com/gitlab-com/www-gitlab-com/blob/master/data/team.yml}.


\captionsetup[figure]{labelfont={bf,small},textfont={it,small}}
\captionsetup[table]{labelfont={bf,small},textfont={it,small}}
\captionsetup[subfloat]{labelfont={bf,small},textfont={it,small}, subrefformat=parens}

\begin{figure*}[t]
    \centering
    \subfloat[]{\label{fig:commit-raw}{\includegraphics[width=0.4\linewidth]{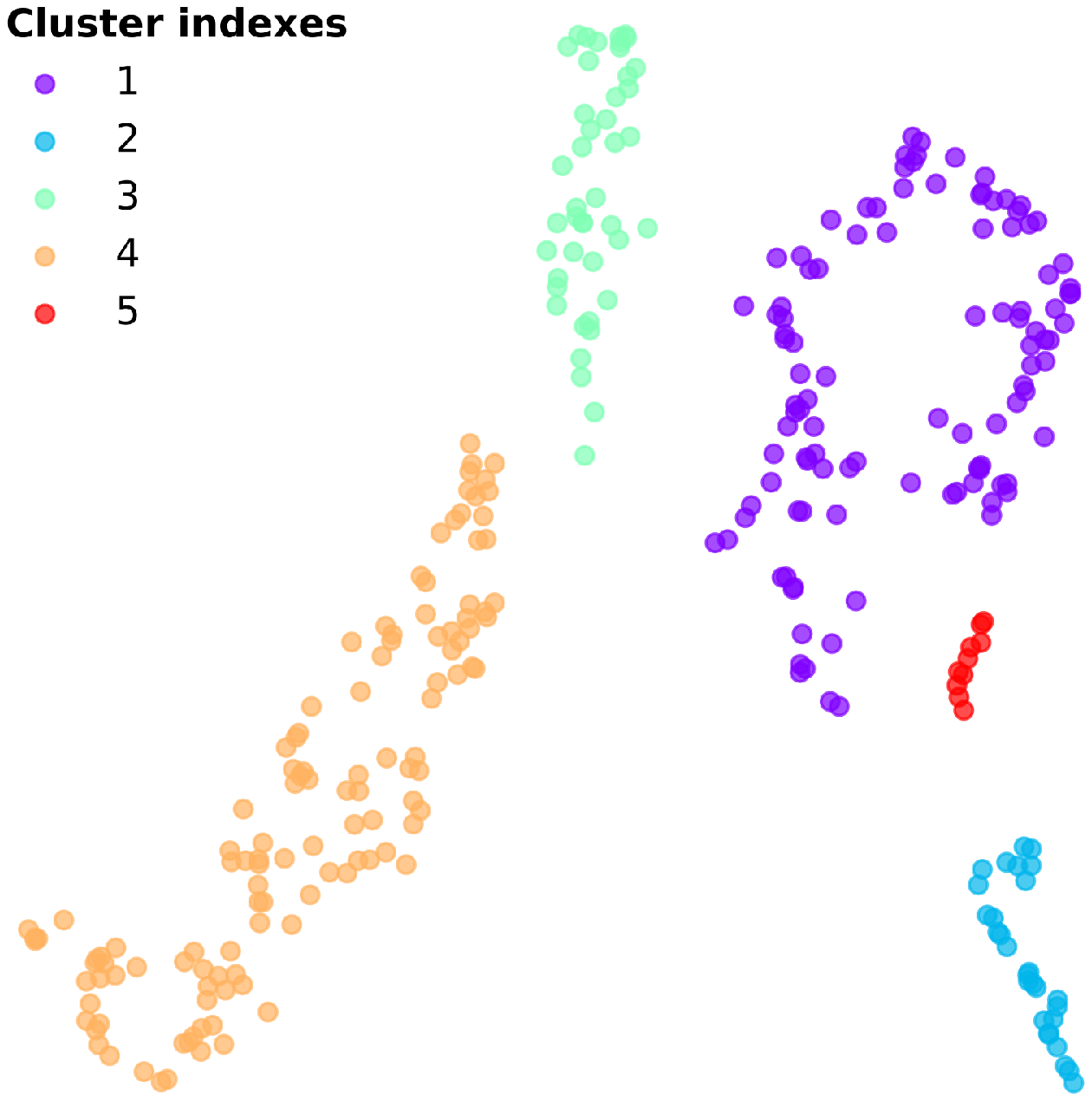}}}\hspace*{3em}
    \subfloat[]{\label{fig:commit-teams}{\includegraphics[width=0.46\linewidth]{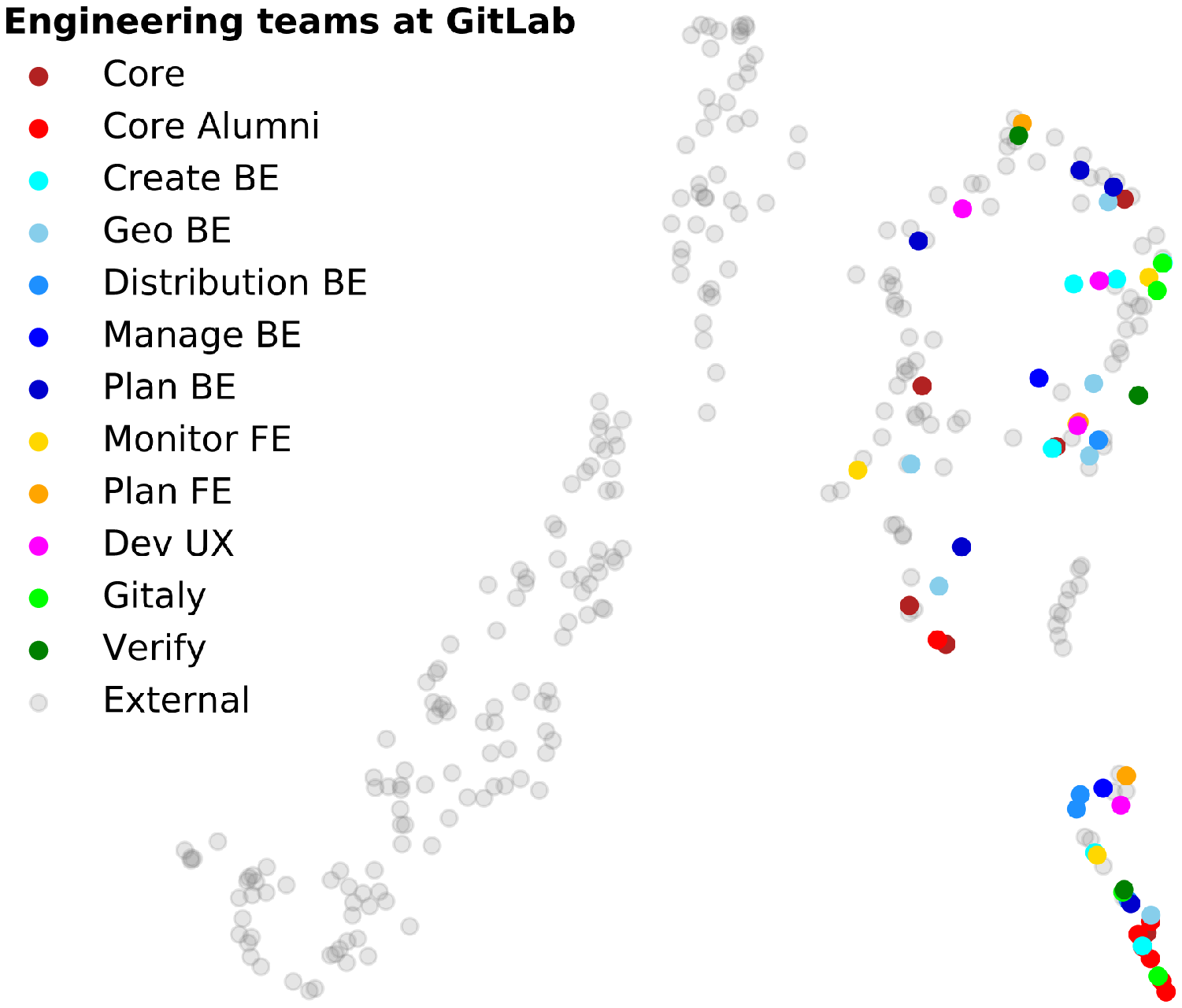}}}\\
    \hspace*{-1.8em}\subfloat[]{\label{fig:experience-raw}{\includegraphics[width=0.46\linewidth]{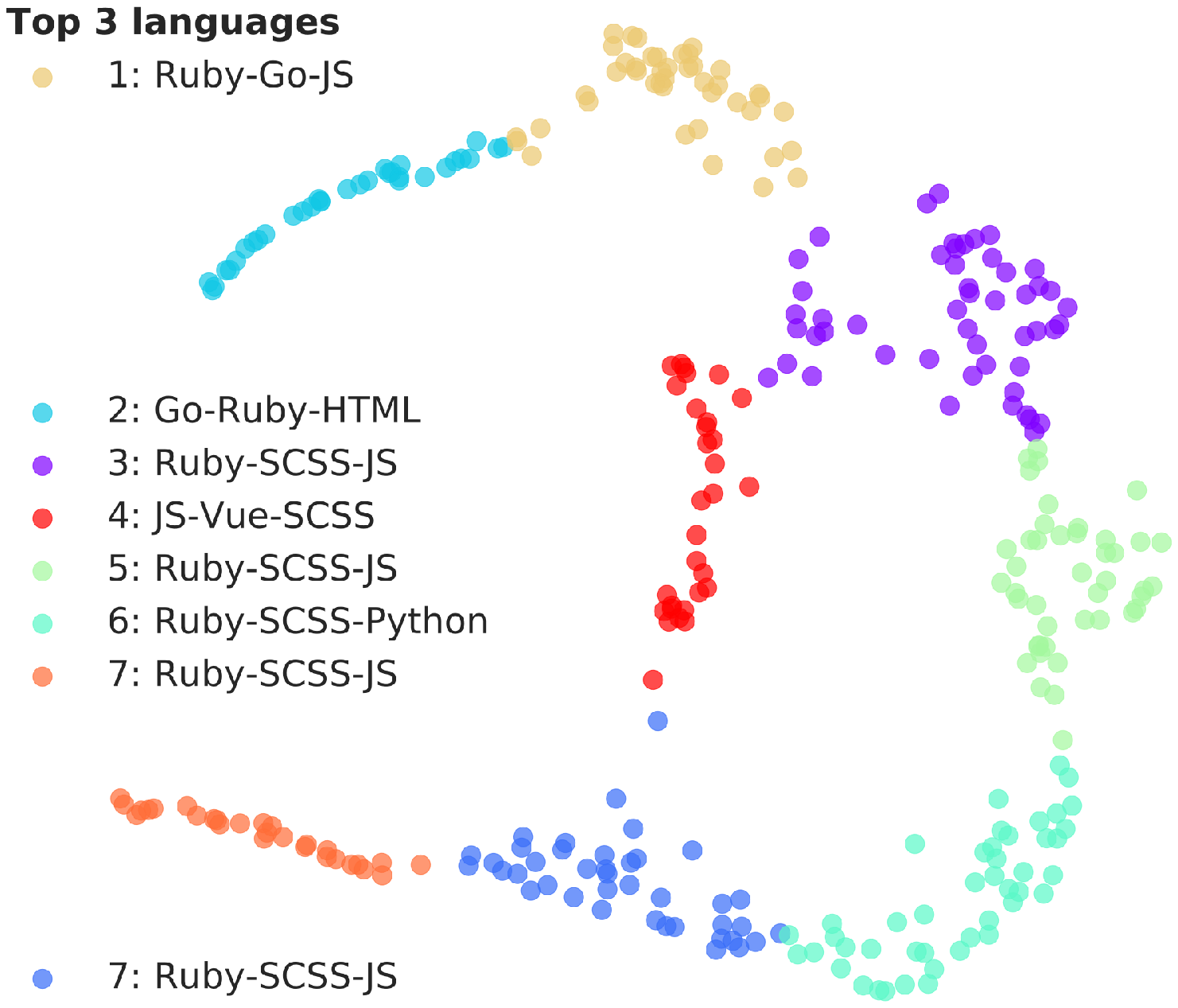}}}\hspace*{3em}
    \subfloat[]{\label{fig:experience-teams}{\includegraphics[width=0.4\linewidth]{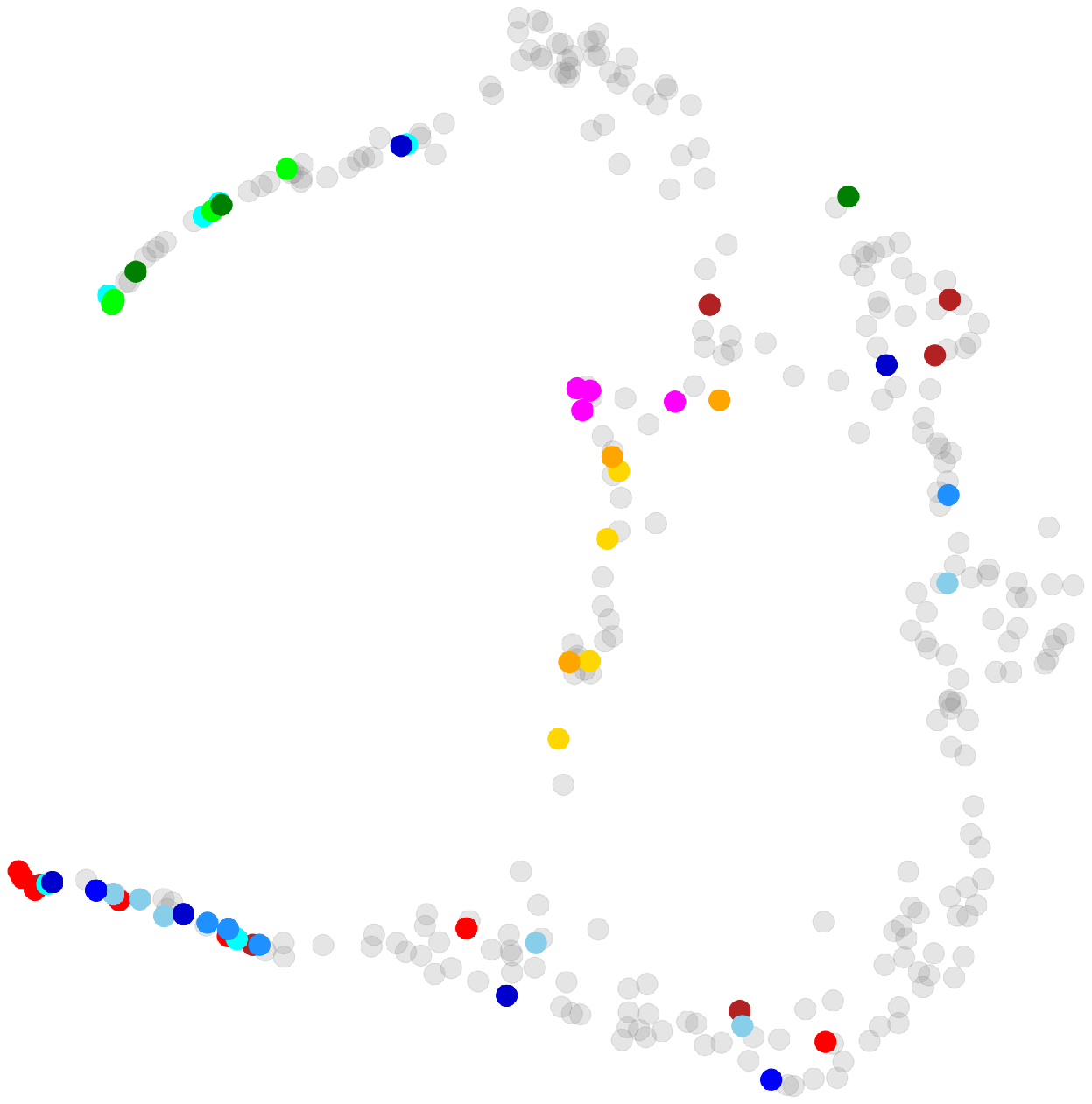}}}
    \caption{Clusters of the contributors to \texttt{gitlab-org} open-source projects based on (a) commit time series and (c) contributed LoC by language. In (b) and (d) the same contributors are labeled not according to their cluster but according to their official team in the \Gitlab{} organization. The gray points stand for contributors that are either outside the company as of April 2019 or using different identities.}
    \label{fig:experience-and-commit}
\end{figure*}

Fig.~\ref{fig:commit-raw} and Fig.~\ref{fig:commit-teams} reveal the temporal commit coupling of the contributors. \Gitlab{} employees are sorted into two clusters among five; commit activity alignment proves to be a reliable way to tell \Gitlab{} employees from external contributors. However, the finer structure often does not respect the teams. For example, cluster \#2 groups high frequency committers across the company, one of them being GitLab's CEO.

On the other hand, clustering developers by their experience in programming languages--- Fig.~\ref{fig:experience-raw} and~\ref{fig:experience-teams}---provides different insights.
Fig.~\ref{fig:experience-teams} demonstrates that people in the same team tend to have similar experience.
All the clusters but \#4 in Fig.~\ref{fig:experience-raw} contain Ruby developers. Indeed, \Gitlab{} is driven by Ruby on Rails and being proficient in Ruby is essential for the company.
Cluster \#2 joins developers who code in both Go and Ruby, which agrees with the current partial conversion of the \Gitlab{} backend from Ruby on Rails to Go\footnote{https://about.gitlab.com/2018/10/29/why-we-use-rails-to-build-gitlab/}. Fig.~\ref{fig:experience-teams} suggests that the \emph{Gitaly}, \emph{Create BE} and \emph{Verify} teams are particularly working on this, since their members are located together.
Then, cluster \#4 emphasizes the recent switch to Vue.js in the frontend, and the \emph{Dev UX} team seems to be one of the team in charge of this. Cluster \#4 clearly gathers all the frontend engineers who code in JavaScript, Vue.js framework, and write SCSS.
Overall, the clusters match the typical Ruby on Rails relation between frontend and backend technologies.

\begin{table}[htb]
    \begin{minipage}{.49\textwidth}
        {\scriptsize
            \captionof{table}{Teams of topic contributors.}
            \label{tab:topics-teams}
            \begin{tabular}{lll}
                \toprule
                Frontend & Config Management & Gollum (wiki) \\
                \midrule
                Verify FE & Monitor BE & Product Management \\
                Core & Plan BE & Support Department \\
                Manage FE & Core Alumni & Core Alumni \\
                Create FE & Create BE & Core \\
                Monitor FE & Distribution BE \\
                Serverless FE & & \\
                Frontend & & \\
                \bottomrule
            \end{tabular}
        }
    \end{minipage}
    \hspace*{1.2em}
    \begin{minipage}{.5\textwidth}
        {\scriptsize
            \captionof{table}{Manual labels assigned to topics.}
            \label{tab:topics-terms}
            \begin{tabular}{llll}
                \toprule
                \multicolumn{1}{c}{Topic label} & \multicolumn{3}{c}{Top terms} \\
                \midrule
                Backend frameworks & servlet & flask & javax \\
                Language detection & languag & java & linguist \\
                Data mining & chartj & graphql & averag \\
                Frontend + UI, CSS & modernizr & mstyle & elementn \\
                Config management & chef & runner & platform \\
                TODO & todo & todos & app \\
                Low-level backend & btree & opclass & using \\
                \bottomrule
            \end{tabular}
        }
    \end{minipage}
\end{table}

Finally, to evaluate the quality of a code topic, we retrieve the teams of the contributors for whom the evaluated topic is the preponderant one. Results shown in Table~\ref{tab:topics-teams} highlight that the topic modeling analysis strongly correlates with \Gitlab{} teams. Table~\ref{tab:topics-terms} shows examples of manually created labels.

We also measure the agreement between topics and the clusters from Fig.~\ref{fig:commit-raw} and \ref{fig:experience-raw}. On average, developers that have a given main topic are in 1.54 and 1.63 clusters for commit activity and experience clustering respectively (the expectations are 2.11 and 2.27 respectively for random clusters).

\section{Conclusion}
\label{sec:conclusion}

This paper demonstrated how to identify existing collaborations in an organization by exploring the topological structure of three feature spaces of a codebase: commit activity, usage of programming languages, and topics of source code identifiers. Clustering proved insightful: we observed a good match with existing development teams in the \Gitlab{} organization, and additionally found several product-oriented inter-organizational squads. 
We traced the partial transition from Ruby to Go in \Gitlab{}'s backends, which was indeed decided at the company level. Commit series alignment distinguished between external and internal contributors well.

Our future research directions include leveraging the commits to external repositories since developers usually contribute to several open source projects that do not belong to their company.
Furthermore, it would be interesting to study how engineering teams have evolved as the company has grown. We believe that combining the recent advancements in machine learning with the expressive features can shed light on relevant coding collaboration opportunities.


\bibliographystyle{ml4se2019}
{\scriptsize\bibliography{main}}

\end{document}

%% file: math_commands.tex
\usepackage{amsmath,amsfonts,bm}

\def\eqref#1{equation~\ref{#1}}

\def\1{\bm{1}}

\DeclareMathAlphabet{\mathsfit}{\encodingdefault}{\sfdefault}{m}{sl}
\SetMathAlphabet{\mathsfit}{bold}{\encodingdefault}{\sfdefault}{bx}{n}

